\documentclass[12pt]{article}
\usepackage{amsfonts}

\usepackage{graphicx}
\usepackage{amsmath}
\usepackage{float}

\textwidth=6.5in \hoffset=-.75in \voffset=-1in \numberwithin{equation}{section}
\numberwithin{figure}{section}

\begin{document}

\begin{titlepage}
\vspace{1cm}
\begin{center}
{\Large \bf {Cosmological Solutions on Atiyah-Hitchin Space in Five Dimensional Einstein-Maxwell-Chern-Simons Theory}}\\
\end{center}
\vspace{2cm}
\begin{center}
A. M. Ghezelbash{ \footnote{E-Mail: masoud.ghezelbash@usask.ca}}
\\
Department of Physics and Engineering Physics, \\ University of Saskatchewan, \\
Saskatoon, Saskatchewan S7N 5E2, Canada\\
\vspace{1cm}

\end{center}

\begin{abstract}
We construct non-stationary exact solutions to five dimensional Einstein-Maxwell-Chern-Simons theory with positive cosmological constant. The solutions are based on four-dimensional Atiyah-Hitchin space.
In asymptotic regions, the solutions approach to Gibbons-Perry-Sorkin monopole solutions. On the other hand, near the four-dimensional bolt of Atiyah-Hitchin space, our solutions show a bolt structure in five dimensions. The c-function for the solutions shows monotonic increase in time, in agreement with the general expected behaviour of c-function in asymptotically dS spacetimes. 
 
\end{abstract}
\end{titlepage}\onecolumn 
\bigskip 

\section{Introduction}

The Einstein-Maxwell-(Dilaton-(Axion)) or -(Chern-Simons) theories in different dimensionalities have been extensively explored from many different directions. The black hole solutions have been considered in \cite{EY1} as well as solitonic and gravitational instantons, dyonic and pp-wave solutions in \cite{EY2}, supergravity solutions in \cite{EY3}, brane worlds and cosmology in \cite{EY4}, NUT and Bolt solutions, Liouville potential, rotating solutions and string theory extensions of Einstein-Maxwell fields in \cite{EY5}.
The Atiyah-Hitchin space is a part of the set of two monopole solutions of
Bogomol'nyi equation. The moduli space of solutions is a product of
$\mathbb{R}^{3}\otimes S^{1}$ (that describes the center of mass
of two monopoles and a phase factor that is related to the total electric
charge of the system) and a four
dimensional manifold $M$, which has self-dual curvature. The
self-duality comes from the hyper-K\"{a}hler property of the moduli space.
A further aspect
concerning $M$ is that it should be rotational invariant, since two
monopoles do exist in ordinary flat space; hence the metric on $M$
can be expressed in terms of three functions of the monopole separation.
Self-duality implies that these three functions obey a set of first-order
ordinary differential equations.
This space has been used recently for construction of five-dimensional three-charge supergravity solutions that only have a rotational $U(1)$
isometry \cite{B4} as well as construction of M-brane solutions \cite{GH}. 
Moreover,  Atiyah-Hitchin space and its various generalizations were identified
with the full quantum moduli space of $\mathcal{N}=4$ supersymmetric gauge
theories in three dimensions \cite{seib}.
Moreover, in the context of string theory and brane world, investigations on 
black hole (ring) solutions in higher dimensions have attracted a lot of attention.
It is believed that in the strong coupling limit, many horizonless
three-charge brane configurations undergo a geometric transition and become
smooth horizonless geometries with black hole or black ring charges \cite{B1}. These charges come completely from fluxes wrapping on non-trivial cycles.
The three-charge black hole (ring) systems are dual to the states of
corresponding conformal field theories: in favor of the idea that non-fundamental-black hole
(ring) systems effectively arise as a result of many horizonless
configurations \cite{B2,Ma1}. 
In eleven-dimensional supergravity, there are solutions based on
transverse four-dimensional hyper-K\"{a}hler metrics (which are equivalent to metrics
with self-dual curvatures). The hyper-K\"{a}hlericity of transverse metric guarantees 
(at least partially) to have supersymmetry \cite{G1}. There are also
many solutions to five-dimensional minimal supergravity. In
five-dimensions, unlike the four dimensions that the only horizon topology
is 2-sphere, we can have different more interesting horizon topologies such
as black holes with horizon topology of 3-sphere \cite{Myers}, black rings
with horizon topology of 2-sphere $\times $ circle \cite{Em1,Em2}, black
saturn: a spherical black hole\ surrounded by a black ring \cite{El1}, black
lens which the horizon geometry is a Lens space $L(p,q)$ \cite{Ch1}. All
allowed horizon topologies have been classified in \cite{Ca1,He1,Ga1}.
Recently, it was shown how a uniqueness theorem might be proved for black
holes in five dimensions \cite{Ho1,Ho2}. It was shown stationary,
asymptotically flat vacuum black holes with two commuting axial symmetries
are uniquely determined by their mass, angular momentum and rod structure.
Specifically, the rod structure \cite{Ha1} determines the topology of
horizon in five dimensions.
In references \cite{Ishi1, Ishi2, Ishi3, Ishi4, Gh3}, the authors constructed 
different solutions of  five-dimensional Einstein-Maxwell theory based on four-dimensional self-dual metrics. Some of solutions describe (multi) black hole solutions with and without cosmological constant. 

Motivated by these facts, in this article we try to construct solutions to five-dimensional Einstein-Maxwell-Chern-Simons theory with positive cosmological constant, based on Atiyah-Hitchin space.
To our knowledge, these solutions are the first solutions to Einstein-Maxwell-Chern-Simons theory with positive cosmological constant that the base space  can't be written in Gibbons-Hawking form.  Moreover, our five dimensional solutions interpolate between a bolt structure to Gibbons-Perry-Sorkin monopole solutions on the boundary. 

We note that hyper-K\"{a}hler Atiyah-Hitchin geometries (unlike Gibbons-Hawking geometries) do not have any tri-holomorphic $U(1)$ isometry, hence our solutions could be used to study the physical processes in spacetimes with positive cosmological constant that do not respect any tri-holomorphic $U(1)$ symmetry.

The outline of this paper is as follows. In section \ref{sec:5Dreview},
we review briefly the Einstein-Maxwell-Chern-Simons theory with positive cosmological constant, the Atiyah-Hitchin space and its features. In section \ref{sec:sol},
we present our cosmological solutions based on two forms for the Atiyah-Hitchin space and discuss the 
asymptotics of the solutions. We conclude in section \ref{sec:con} with a summary of our solutions and possible future research directions.

\section{Five-dimensional Einstein-Maxwell-Chern-Simons Theory with Positive Cosmological Constant and Atiyah-Hitchin Space}

\label{sec:5Dreview}

We consider the five dimensional Einstein-Maxwell-Chern-Simons theory with a positive cosmological constant. The action is given by 
\begin{equation}
S=\frac{1}{16\pi}\int d^5x\, \sqrt{-g}(R-4\Lambda -F_{\mu\nu}F^{\mu\nu}-\frac{2}{3\sqrt{3}}\epsilon^{\mu\nu\rho\eta\xi}F_{\mu\nu}F_{\rho\eta}A_\xi),
\end{equation}
where $R$ and  $F_{\mu\nu}$ are five dimensional Ricci scalar and Maxwell filed.
The Einstein and Maxwell equations are
\begin{eqnarray}
R_{\mu\nu}-(\frac{1}{2}R-2\Lambda) g_{\mu\nu}&=&2(F_{\mu\lambda}F_{\nu}^{\lambda}-\frac{1}{4}g_{\mu\nu}F^2),\label{EEQ} \\
F^{\mu\nu}_{;\nu}&=&\frac{2}{3\sqrt{3}}\epsilon^{\mu\nu\rho\eta\xi}F_{\nu\rho}F_{\eta\xi},\label{GEQ}
\end{eqnarray}
respectively. We take the following form for the five-dimensional metric
\begin{equation}
ds_5^{2}=-H(r,t)^{-2}dt^{2}+H(r,t)ds_{AH}^2,
\label{ds5}
\end{equation}%
and the only non-vanishing component of gauge field as
\begin{equation}
A_t=\frac{\eta\sqrt{3}}{2}\frac{1}{H(r,t)},
\label{gauge5}
\end{equation}%
where $\eta=+1$ or $\eta=-1$. The Atiyah-Hitchin metric $ds_{AH}^{2}$ is given by the following manifestly 
$SO(3)$ invariant form \cite{GM} 
\begin{equation}
ds_{AH}^{2}=f^{2}(r)dr^{2}+a^{2}(r)\sigma _{1}^{2}+b^{2}(r)\sigma
_{2}^{2}+c^{2}(r)\sigma _{3}^{2},  \label{AHmetric}
\end{equation}%
where $\sigma _{i\text{ }}$ are Maurer-Cartan one-forms (see appendix).
The metric (\ref%
{AHmetric}) satisfies Einstein's equations provided%
\begin{eqnarray}
a^{\prime }&=&f\frac{(b-c)^{2}-a^{2}}{2bc}, \label{conditions1}\\ 
b^{\prime }&=&f\frac{(c-a)^{2}-b^{2}}{2ca}, \label{conditions2}\\ 
c^{\prime }&=&f\frac{(a-b)^{2}-c^{2}}{2ab}. \label{conditions3}
\end{eqnarray}
Choosing\textbf{\ }$f(r)=-\frac{b(r)}{r}$\textbf{\ }, we can find the explicit expressions
for the metric functions $a,b$ and $c$ in terms of Elliptic integrals (see appendix). 

In fact as $r\rightarrow \infty ,$ the metric (\ref{AHmetric}) reduces to 
\begin{equation}
ds_{AH}^{2}\rightarrow (1-\frac{2n}{r})(dr^{2}+r^{2}d\theta ^{2}+r^{2}\sin
^{2}\theta d\phi ^{2})+4n^{2}(1-\frac{2n}{r})^{-1}(d\psi +\cos \theta d\phi
)^{2},  \label{reducedAH}
\end{equation}%
which is the well known Euclidean Taub-NUT metric with a negative NUT charge 
$N=-n.$ 
The metric (\ref{reducedAH}) is obtained from a consideration of the limiting behaviors
of the functions $a,b$ and $c$ at large monopole separation (see appendix).
On the other hand, near the bolt location, $r\rightarrow n\pi$, we can write the metric 
(\ref{AHmetric}) as%
\begin{equation}
ds^{2}=d\epsilon^{2}+4\epsilon^{2}(d\widetilde{\psi }+\cos \widetilde{\theta }d%
\widetilde{\phi })^{2}+\pi ^{2}n^{2}(d\widetilde{\theta }+\sin ^{2}%
\widetilde{\theta }d\widetilde{\phi }),  \label{nearbolt}
\end{equation}%
where we use a new set of coordinates $\widetilde{\psi },\widetilde{\theta }$ and $\widetilde{\phi }$, related to $\psi ,\theta ,$\ $\phi $ through some rotations.
We note that the last term in (\ref{nearbolt}) is the induced metric
on the two dimensional bolt.

\section{Cosmological Einstein-Maxwell-Chern-Simons Solutions over Atiyah-Hitchin Base Space}
\label{sec:sol}
To find the five-dimensional metric function $H(r,t)$, we consider equations of motion (\ref{EEQ}-\ref{GEQ}). 
The metric (\ref{ds5}) and the gauge field (\ref{gauge5}) are solutions to Einstein-Maxwell equations with positive cosmological constant, provided $H(r,t)$ is a solution to the
differential equation,
\begin{eqnarray}
& &r^2a(r)c(r)\frac{\partial^2H(r,t)}{\partial r^2}+r(a(r)b(r)+a(r)c(r)+b(r)c(r)-b(r)^2)\frac{\partial H(r,t)}{\partial r}\nonumber \\
&+&\frac{4}{3}
\Lambda H(r,t)^2 a(r)b^2(r)c(r)-H(r,t)^2(\frac{\partial H(r,t)}{\partial t})^2 a(r)b^2(r)c(r)=0.
\label{eqforH}
\end{eqnarray}
The solution to this partial differential equation is given by
\begin{equation}
H(r,t)=\tilde H(r)+2\sqrt{\Lambda /3}t,
\label{HHH}
\end{equation}
where the function $\tilde H(r)$ is
\begin{equation}
\tilde H(r)=H_0+H_1\int dr e^{\int \frac{b(r)^2-a(r)b(r)-a(r)c(r)-b(r)c(r)}{ra(r)c(r)} dr},
\label{Hsol}
\end{equation}
with $H_0$ and $H_1$, two constants of integration. Although the $r$-dependences of metric functions $a,b,c$, are given explicitly in 
equations (\ref{abc1}-\ref{abc3}), it's unlikely to find an analytic expression
for $\tilde H(r)$ given by (\ref{Hsol}). 
As $r\rightarrow \infty$, the metric function (\ref{Hsol}) goes to
\begin{equation}
H(r,t)=2\sqrt{\Lambda /3}t+H_0-\frac{H_1}{r},
\label{Hatinf}
\end{equation}
On the other hand, near bolt, the metric function $\tilde H(r)$ has a logarithmic divergence as 
\begin{equation}
\tilde H(r)\simeq \frac{H_1}{4n^2\pi^2}\ln(\epsilon)+H_0 + O(\epsilon),
\label{Honbolt}
\end{equation}
where $\epsilon=r-n\pi$. This type of divergence in the metric function has been observed previously in the metric function of M2-brane in a transverse Atiyah-Hitchin space \cite{GH}. We note that setting $\Lambda=0$ in (\ref{HHH}) yields the solutions of Einstein-Maxwell equations with no cosmological constant \cite{Gh3}.

\begin{figure}[tbp]
\centering           
\begin{minipage}[c]{.3\textwidth}
        \centering
        \includegraphics[width=\textwidth]{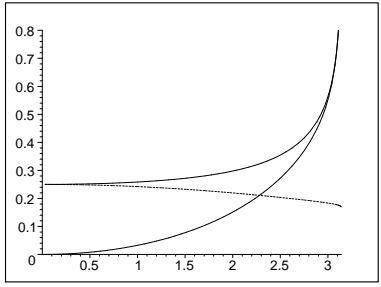}
    \end{minipage}
\caption{
The functions $a$ (solid bottom), $b$ (solid top) and $c$ (dashed) plotted as functions of $\vartheta$. 
}
\label{abcs}
\end{figure}

As we noticed, we could not find a closed analytic expression for the metric function given 
in (\ref{Hsol}). 
To overcome this problem, we choose $f(\xi)$ in (\ref{AHmetric}) to be
$4a(\xi)b(\xi)c(\xi)$ and so the Atiyah-Hitchin metric reads
\begin{equation}
ds_{AH}^{2}=16a^{2}(\xi)b^{2}(\xi)c^{2}(\xi)d\xi^{2}+a^{2}(\xi)\sigma _{1}^{2}+b^{2}(\xi)\sigma
_{2}^{2}+c^{2}(\xi)\sigma _{3}^{2},  \label{AHmetric2}
\end{equation}
where functions $a(\xi),b(\xi)$ and $c(\xi)$ satisfy equations (\ref{conditions1}-\ref{conditions3}) with $f(\xi)=4a(\xi)b(\xi)c(\xi)$
and $'$ means $\frac{d}{d\xi}$. By introducing the new functions $\psi_1(\xi),\psi_2(\xi)$ and $\psi_3(\xi)$ such that 
\begin{eqnarray}
a^2(\xi)&=&\frac{\psi_2\psi_3}{4\psi_1}\label{psis1},\\
b^2(\xi)&=&\frac{\psi_3\psi_1}{4\psi_2}\label{psis2},\\
c^2(\xi)&=&\frac{\psi_1\psi_2}{4\psi_3}\label{psis3},
\end{eqnarray}
the set of equations (\ref{conditions1}-\ref{conditions3}) with $f(\xi)=4a(\xi)b(\xi)c(\xi)$ reduces onto a Darboux-Halpern system
\begin{eqnarray}
\frac{d}{d\xi}(\psi_1+\psi_2)+2\psi_1\psi_2&=&0,\label{H1}\\
\frac{d}{d\xi}(\psi_2+\psi_3)+2\psi_2\psi_3&=&0,\label{H2}\\
\frac{d}{d\xi}(\psi_3+\psi_1)+2\psi_3\psi_1&=&0.\label{H3}
\end{eqnarray} 
We can find the solutions to the above equations as 
\begin{eqnarray}
\psi_1&=&-\frac{1}{2}(\frac{d}{d\vartheta}\mu ^2+\frac{\mu ^2}{\sin\vartheta}),
\label{psi1} \\
\psi_2&=&-\frac{1}{2}(\frac{d}{d\vartheta}\mu ^2-\frac{\mu ^2\cos\vartheta }{\sin\vartheta}),
\label{psi2} \\
\psi_3&=&-\frac{1}{2}(\frac{d}{d\vartheta}\mu ^2-\frac{\mu ^2}{\sin\vartheta}),
\label{psi3} 
\end{eqnarray}
where 
\begin{equation}
\mu (\vartheta)=\frac{1}{\pi}\sqrt{\sin\vartheta}K(\sin\frac{\vartheta}{2}). \label{varw}
\end{equation}
The new coordinate $\vartheta$ is related to the coordinate $\xi$ by 
\begin{equation}
\xi=-\int _\vartheta ^ \pi \frac{d\vartheta}{\mu ^2 (\vartheta)}. 
\label{theta}
\end{equation}
The coordinate $\vartheta$ is a monotonic increasing function of $\xi$; takes values over $[0,\pi]$ if 
the coordinate $\xi$ is chosen to take values on $(-\infty,0]$. 
The function $\mu (\vartheta)$ has an increasing behavior from $\vartheta=0$ to $\vartheta_0=2.281318$. At 
$\vartheta=\vartheta_0$, the
function $\mu$ reaches to maximum value $0.643243$ and decreases then to zero at $\vartheta=\pi$. Hence, in the range of $0 < \vartheta <\pi$, $\mu$ is positive and so
the change of variables, given in (\ref{theta}), is completely well defined.
The functions $\psi_1,\psi_2$ are always negative and $\psi_3$ is always positive, hence, equations (\ref{psis1}, \ref{psis2}, \ref{psis3}) show the metric functions always are positive. In figure (\ref{abcs}), the behaviors of functions $a,b$ and $c$ versus $\vartheta$ are plotted.

The five-dimensional metric and gauge field are given by
\begin{equation}
ds^2=-\frac{dt^2}{(\alpha\xi+\beta+\lambda t)^2}+(\alpha\xi+\beta+\lambda t)\{16a^{2}(\xi)b^{2}(\xi)c^{2}(\xi)d\xi^{2}+a^{2}(\xi)\sigma _{1}^{2}+b^{2}(\xi)\sigma
_{2}^{2}+c^{2}(\xi)\sigma _{3}^{2}\},\label{AH5ex}
\end{equation}
and
\begin{equation}
A_t=\frac{\eta\sqrt{3}}{2(\alpha\xi+\beta+\lambda t)}.\label{gaugefin}
\end{equation}
where $\alpha$ and $\beta$ are two constants and $\lambda=2\sqrt{\Lambda/3}$. In asymptotic region where $\xi \rightarrow 0$ (or $\vartheta=\pi-\epsilon$), the metric (\ref{AH5ex}) reduces to
\begin{equation}
ds^2=-d\tau^2+\frac{e^{2\lambda\tau}}{4\pi^2}\{d\zeta^2+\zeta^2d\Omega^2+(d\psi+\cos\theta d\phi)^2\}
\label{asymp}
\end{equation}
where $\tau=\frac{\ln(\beta+\lambda t)}{\lambda}$
and the coordinate $\zeta$ is related to $\epsilon$ by $\zeta=-\ln\epsilon$ and to $\xi$ by $\zeta=-\frac{\pi^2}{\xi}$. We choose $\beta \geq 1$ and we have $0 \leq \lambda\tau < +\infty$ by considering the positive time coordinate in the metric (\ref{AH5ex}).  
At a fixed $\tau$ slice, the metric (\ref{asymp}) represents an Euclidean Taub-Nut metric with a negative NUT charge $N=-\frac{1}{2\pi}$. The Ricci scalar of the spacetime (\ref{asymp}) is equal to $20\lambda^2-2\pi^2e^{-2\lambda\tau}/\zeta^4$ that approaches to $20\lambda^2$ as $\zeta \rightarrow -\infty$. In this limit, the Kretschman invariant approaches $40\lambda^4+O(1/\zeta ^6)$.
We note the asymptotic metric (\ref{asymp}) represents the Gross-Perry-Sorkin monopole solution with a cosmological constant. The Killing vectors of Gross-Perry-Sorkin monopole solution (\ref{asymp}) are
\begin{eqnarray}
\xi_1&=&\frac{\partial}{\partial \psi}\\
\xi_2&=&\frac{\partial}{\partial \phi}\\
\xi_3&=&\frac{\sin\phi}{\sin\theta}
\frac{\partial}{\partial \psi}
+\cos\phi\frac{\partial}{\partial \theta}
-\frac{\cos\theta\sin\phi}{\sin\theta}\frac{\partial}{\partial \phi}\\
\xi_4&=&\frac{\cos\phi}{\sin\theta}
\frac{\partial}{\partial \psi}
-\sin\phi\frac{\partial}{\partial \theta}
-\frac{\cos\theta\cos\phi}{\sin\theta}\frac{\partial}{\partial \phi}
\end{eqnarray}
and all of them are space-like. So, we conclude our solution (\ref{AH5ex}) is a non-stationary solution.  
There is a singualrity at $t=0$ which can be avoided by choosing sign of $\alpha$ opposite to that of $\beta$. Moreover, to avoid any singularity at $t>0$ and to get a regular positive definite metric, we should choose $\alpha < 0$ and $\beta > 0$. 

On the other extreme limit where $\xi \rightarrow -\infty$, (or $\vartheta \rightarrow 0$), 
we find the metric (\ref{AH5ex}) changes to
\begin{equation}
ds^2=-\frac{dt^2}{\alpha^2\xi^2}+\alpha \xi \{
\frac{d\varrho ^2}{4}+\varrho^2 \sigma_1^2+\frac{1}{16}(\sigma_2^2+\sigma_3^2)\}.
\label{bol2}
\end{equation}
where $\varrho=\frac{e^{\xi /2}}{32}$ (see Appendix).
The metric (\ref{bol2}) shows a bolt at $\varrho=0$ since Ricci scalar and Kretschman invariant of the metric (\ref{bol2}) diverge on bolt location $\varrho=0$. 

As it is well known, in both asymptotically AdS/dS spacetimes, there is a natural correspondence between
phenomena occurring near the boundary (or in the deep interior) of either
spacetime and UV (IR) physics in the dual CFT. Solutions that are
asymptotically (locally) dS lead to an interpretation in terms of
renormalization group flows and an associated generalized dS $c$-theorem.
The theorem states that in a contracting patch of dS spacetime, the
renormalization group flows toward the infrared and in an expanding
spacetime, it flows toward the ultraviolet. 
The $c$-function for representation of the dS metric with a wide
variety of boundary geometries involving direct products of flat space, the
sphere and hyperbolic space was studied in \cite{Leb}. Since our 5-dimensional exact solution 
(\ref{AH5ex}) is asymptotically dS, we consider the five-dimensional 
$c$-function as 

\begin{equation}
c=\frac{1}{\left( G_{\mu \nu }n^{\mu }n^{\nu }\right) ^{\frac{3}{2}}}\label{cfun}
\end{equation}%
where $n^{\mu }$\ is the unit vector in $t$ direction. Consequently $c$-theorem 
implies that $c$-function must increase (decrease) for any expanding (contracting) patch of the
spacetime.

\begin{figure}[tbp]
\centering           
\begin{minipage}[c]{.3\textwidth}
        \centering
        \includegraphics[width=\textwidth]{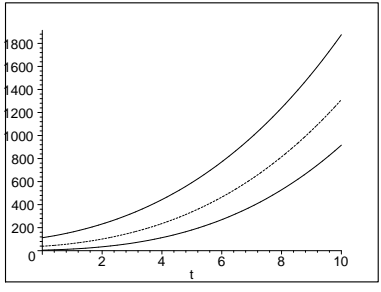}
    \end{minipage}
\caption{The $c$-function of five dimensional spacetime (\ref{AH5ex}) plotted as a function of $%
t$ for three different fixed values of $\xi$. The upper and lower solid curves correspond to $\xi_1$ and $\xi_3$, respectively and the dashed curve to $\xi_2$.}
\label{CvTime}
\end{figure}

\begin{figure}[tbp]
\centering           
\begin{minipage}[c]{.3\textwidth}
        \centering
        \includegraphics[width=\textwidth]{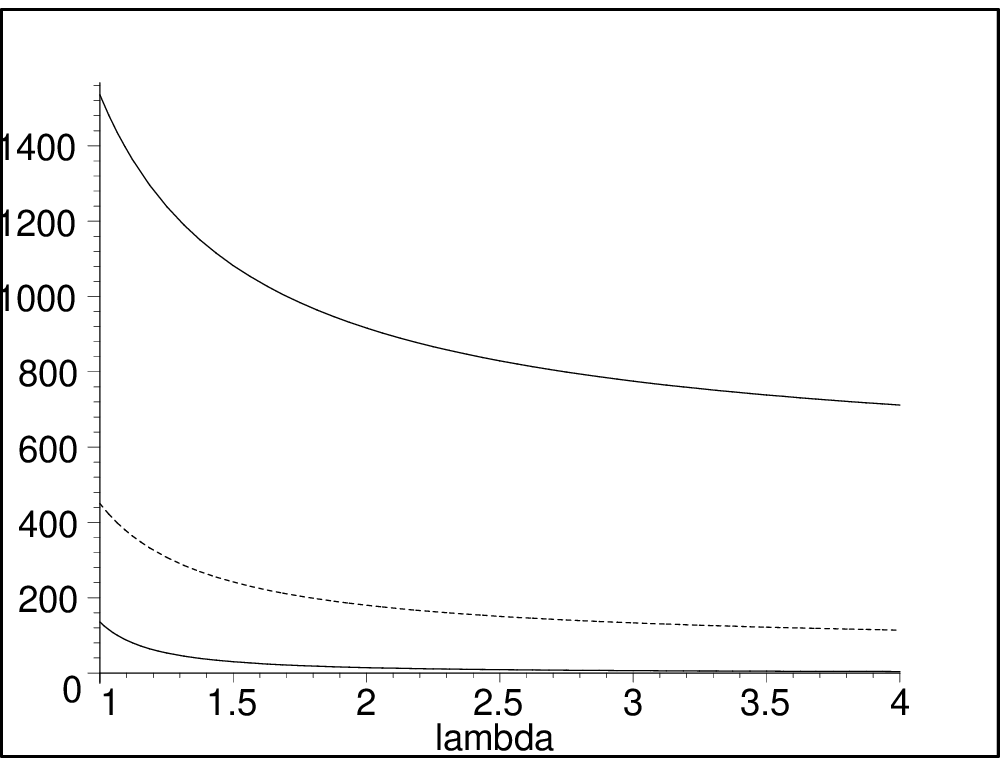}
    \end{minipage}
\caption{The $c$-function of five dimensional spacetime (\ref{AH5ex}) as function of $%
\lambda$ for three different fixed values of $t$. The upper and lower solid curves correspond to $t=1$ and $t=10$, respectively and the dashed curve to $t=5$.}
\label{Cvlambda}
\end{figure}

For our solutions (\ref{AH5ex}), the $c$-function reads as%
\begin{equation}
c(\xi,t)=\frac{8}{9}\sqrt{3}\frac{a(\xi)^{3/2}b(\xi)^3c(\xi)^{3/2}\{\alpha\xi+\beta+\lambda t\}^{15/2}}{X(\xi,t)^{3/2}}
\label{cfunKdS}
\end{equation}%
where
\begin{eqnarray}
X(\xi,t)&=&2r\alpha b(\xi)^2\beta+2\xi^2\alpha^2b(\xi)^2-2c(\xi)\xi b(\xi)\alpha \lambda t+2\xi\alpha b(\xi)^2\lambda t-2c(\xi)\xi^2b(\xi)\alpha^2-
\nonumber\\
&-&2c(\xi)\xi b(\xi)\alpha\beta-2\xi^2a(\xi)\alpha^2 b(\xi)-2\xi a(\xi)\alpha b(\xi)\beta-2\xi a(\xi)\alpha b(\xi)\lambda t-
\nonumber\\
&-&\xi^2\alpha^2 a(\xi)c(\xi)-2\xi\alpha a(\xi)c(\xi)\beta-2\xi\alpha a(\xi)c(\xi)\lambda t+2\lambda^2 b(\xi)^2a(\xi)c(\xi)\alpha^3\xi^3+
\nonumber\\
&+&6\lambda^2b(\xi)^2a(\xi)c(\xi)\alpha^2\xi^2\beta+6\lambda^2 b(\xi)^2a(\xi)c(\xi)\alpha \xi\beta^2+6\lambda^3b(\xi)^2a(\xi)c(\xi)\alpha^2\xi^2t+
\nonumber\\
&+&12\lambda^3b(\xi)^2a(\xi)c(\xi)\alpha \xi\beta t+2\lambda^2b(\xi)^2a(\xi)c(\xi)\beta^3+6\lambda^3b(\xi)^2a(\xi)c(\xi)\beta^2 t+
\nonumber\\
&+&6\lambda^4b(\xi)^2a(\xi)c(\xi)\alpha \xi t^2+6\lambda^4b(\xi)^2a(\xi)c(\xi)\beta t^2+2\lambda^5b(\xi)^2a(\xi)c(\xi)t^3
\end{eqnarray}
We notice the $c$-function depends
explicitly on time as well as coordinate $\xi$. To  understand the behaviour of $c$-function versus time, we consider three different values for $\vartheta$ to be equal to $1,2,3$. The corresponding $\xi$'s are equal to $\xi_1=-7.513419, \xi_2=-4.430164$ and $\xi_3=-1.734927$. In figure (\ref{CvTime}), the $c$%
-functions for three different fixed values of $\xi$ are plotted where we set $\alpha=-1,\beta=2$ and the cosmological constant $\Lambda =3$.
As one can see from figure (\ref{CvTime}), the $c$
-function is a monotonically increasing function of $t$ for any value of $\xi$ which shows
expansion of a constant $t$-surface of the metric (\ref{AH5ex}).
In fact, we notice our solutions (\ref{AH5ex}) at future infinity ($t \rightarrow +\infty$) approach to
\begin{equation}
ds^2=-dT^2+\lambda e^{\lambda T}ds^2_{AH}
\label{fi}
\end{equation} 
where $T=\frac{\ln t}{\lambda}$. Hence the scale factor in (\ref{fi}), expands exponentially near future infinity. At future infinity, the $c$-function (\ref{cfunKdS}) behaves as $t^3$; in agreement with one expects from $c$-theorem. According to $c$-theorem for any asymptotically locally dS spacetimes, the $c$-function should increase/decrease for any expanding/contracting patch of the spacetime.
Moreover, in figure (\ref{Cvlambda}), the $c$%
-function versus cosmological parameter $\lambda$ for three different fixed values of $t$ is plotted where we set $\alpha=-1,\beta=2$ and $\xi=\xi_3$.

\section{Concluding Remarks}

\label{sec:con}

We construct exact non-stationary cosmological solutions to five-dimensional Einstein-Maxwell-Chern-Simons theory based on four-dimensional Atiyah-Hitchin space.
We consider the Atiyah-Hitchin metric in form of (\ref{AHmetric2}); as a result we get the simple analytic solutions for the metric function of (\ref{ds5}). The five-dimensional metric function is a linear combination of time and the radial variable of four-dimensional Atiyah-Hitchin base space.  The cosmological solutions are regular everywhere except on the location of original bolt in four-dimensional Atiyah-Hitchin base space. The same type of regularity was observed in higher-dimensional (super)gravity solutions based on transverse self-dual hyper-K\"{a}hler manifolds \cite{GH, HC, GH2}.
It's quite possible the irregular hypersurface(s) of our solutions can be converted to regular hypersurface(s) in five-dimensional space-time if we consider more coordinate dependence in metric function. The other open issue is the thermodynamics of constructed solutions in this paper as well as application of dS/CFT correspondence to our solutions. We leave these items for a future article.
We note that the $c$-function for our solutions depends on cosmological time as well as the radial coordinate of Atiyah-Hitchin base space. For any fixed value of radial coordinate, the $c$-function is monotonically increasing with time, in perfect agreement with expansion of $t$-fixed hypersurfaces and $c$-theorem.

\bigskip

{\Large Acknowledgments}

\bigskip

This work was supported by the Natural Sciences and Engineering Research
Council of Canada.

\bigskip

{\Large Appendix}

\bigskip

The Maurer-Cartan one-forms are given by 
\begin{eqnarray}
\sigma _{1}&=&-\sin \psi d\theta +\cos \psi \sin \theta d\phi, \label{mcFORMS1}\\ 
\sigma _{2}&=&\cos \psi d\theta +\sin \psi \sin \theta d\phi, \label{mcFORMS2}\\ 
\sigma _{3}&=&d\psi +\cos \theta d\phi, \label{mcFORMS3}
\end{eqnarray}
with the property 
\begin{equation}
d\sigma _{i}=\frac{1}{2}\varepsilon _{ijk}\sigma _{j}\wedge \sigma _{k}.
\label{dsigma}
\end{equation}%
The metric on the $\mathbb{R}^{4}$ (with a radial coordinate $R$
and Euler angles ($\theta ,\phi ,\psi $) on an $S^{3}$) could be written in
terms of Maurer-Cartan one-forms by 
\begin{equation}
ds^{2}=dR^{2}+\frac{R^{2}}{4}(\sigma _{1}^{2}+\sigma _{2}^{2}+\sigma
_{3}^{2}).  \label{s3METRIC}
\end{equation}%
We also note that $\sigma _{1}^{2}+\sigma _{2}^{2}$ is the standard metric
of the round unit radius $S^{2}$ and $4(\sigma _{1}^{2}+\sigma
_{2}^{2}+\sigma _{3}^{2})$ gives the same for $S^{3}.$
The solutions to differential equations (\ref{conditions1}-\ref{conditions3}) with $f(r)=-\frac{b(r)}{r}$
are given explicitly by
\begin{eqnarray}
a(r)&=&\sqrt{\frac{r\Upsilon \sin (\gamma )\{\frac{1-\cos (\gamma )}{2}r-\sin
(\gamma )\Upsilon \}}{\Upsilon \sin (\gamma )+r\cos ^{2}(\frac{\gamma }{2})}},
\label{abc1}\\ 
b(r)&=&\sqrt{\frac{\{\Upsilon \sin (\gamma )-\frac{1-\cos \gamma }{2}%
r\}r\{-\Upsilon \sin (\gamma )-\frac{1+\cos \gamma }{2}r\}}{\Upsilon \sin
(\gamma )}}, \label{abc2}\\ 
c(r)&=&-\sqrt{\frac{r\Upsilon \sin (\gamma )\{\frac{1+\cos (\gamma )}{2}r+\sin
(\gamma )\Upsilon \}}{-\Upsilon \sin (\gamma )+\frac{1-\cos \gamma }{2}r}},\label{abc3}
\end{eqnarray}%
where 
\begin{equation}
\Upsilon =\frac{2nE\{\sin (\frac{\gamma }{2})\}}{\sin (\gamma )}-\frac{%
nK\{\sin (\frac{\gamma }{2})\}\cos (\frac{\gamma }{2})}{\sin (\frac{\gamma }{%
2})},  \label{GAMMA}
\end{equation}%
and 
\begin{equation}
K(\sin (\frac{\gamma }{2}))=\frac{r}{2n}.  \label{gama}
\end{equation}%
Here $K$ and $E$ are the elliptic integrals 
\begin{eqnarray}
K(k) &=&\int_{0}^{1}\frac{dt}{\sqrt{1-t^{2}}\sqrt{1-k^{2}t^{2}}}%
=\int_{0}^{\pi /2}\frac{d\theta }{\sqrt{1-k^{2}\cos ^{2}\theta }},
\label{Ell} \\
E(k) &=&\int_{0}^{1}\frac{\sqrt{1-k^{2}t^{2}}dt}{\sqrt{1-t^{2}}}%
=\int_{0}^{\pi /2}\sqrt{1-k^{2}\cos ^{2}\theta }d\theta,
\end{eqnarray}%
and the coordinate $r$ ranges over the interval $[n\pi ,\infty )$, which
corresponds to $\gamma \in \lbrack 0,\pi ).$ The positive number $n$\ is a
constant number with unit of length that is related to NUT charge of metric
at infinity obtained from Atiyah-Hitchin metric (\ref{AHmetric}).
At large monopole separation, $r\rightarrow \infty$, the metric functions behave as
\begin{eqnarray}
a(r)&=&r(1-\frac{2n}{r})^{1/2}+O(e^{-r/n}), \label{abcatinfinity1}\\ 
b(r)&=&r(1-\frac{2n}{r})^{1/2}+O(e^{-r/n}), \label{abcatinfinity2}\\ 
c(r)&=&-2n(1-\frac{2n}{r})^{-1/2}+O(e^{-r/n}). \label{abcatinfinity3}%
\end{eqnarray}%
In the other extreme limit where $\epsilon=r-n\pi\rightarrow 0$, from equations (\ref{abc1}-\ref{abc3}), we find the following behaviors for the metric functions $a(r), b(r)$ and $c(r)$ 
\begin{eqnarray}
a(r)&=&2\epsilon+O(\epsilon^2),\label{anp}\\
b(r)&=&n\pi+\frac{\epsilon}{2}+O(\epsilon^2),\label{bnp}\\
c(r)&=&-n\pi+\frac{\epsilon}{2}+O(\epsilon^2).\label{cnp}
\end{eqnarray}
Equation (\ref{anp}) shows clearly a bolt singularity as
$\epsilon \rightarrow 0$.
We consider here the limiting behaviour of Atiyah-Hitchin metric (\ref{AHmetric2}) where the metric functions are given by (\ref{psi1}-\ref{psi3}). In the extreme limit of $\xi \rightarrow -\infty$ (that corresponds to $\vartheta \rightarrow 0$), the metric functions (\ref{psi1}-\ref{psi3}) behave as
\begin{eqnarray}
a&\simeq&\frac{\vartheta ^2}{768}(24+\vartheta^2+O(\vartheta^4)),   \label{aatxiinf}\\
b&\simeq&\frac{1}{4}(1+\frac{\vartheta ^2}{32}+O(\vartheta^4)),   \label{batxiinf}\\
c&\simeq&\frac{1}{4}(1-\frac{\vartheta ^2}{32}+O(\vartheta^4)).   \label{catxiinf}
\end{eqnarray}
In this limit, the Elliptic integral in equation (\ref{varw}) approaches
\begin{equation}
K(\sin\frac{\vartheta}{2})\simeq \frac{\pi}{2}(1+\frac{1}{16}\vartheta^2+
O(\vartheta^4)),
\end{equation}
hence from equations (\ref{varw}) and (\ref{theta}), we get
\begin{equation}
\xi\simeq 4\ln \vartheta, 
\end{equation}
and so the metric (\ref{AH5ex}) changes to
\begin{equation}
ds^2=-\frac{dt^2}{\alpha^2\xi^2}+\alpha \xi \{
\frac{d\varrho ^2}{4}+\varrho^2 \sigma_1^2+\frac{1}{16}(\sigma_2^2+\sigma_3^2)\}.
\label{bol}
\end{equation}
where $\varrho=\frac{e^{\xi /2}}{32}$.

\end{document}